\documentclass[	
						superscriptaddress,%
						twocolumn,%
						a4paper,
						showkeys,
					]{revtex4}

\usepackage[frenchb,english]{babel}
\usepackage[T1]{fontenc}
\usepackage{upgreek}
\usepackage{textcomp} 				
\usepackage[dvips]{graphicx}		
\usepackage{color}					
\usepackage[nice]{nicefrac} 		
\usepackage{amssymb,amsmath}		
\usepackage{array}					

\begin{document}
\title{Thulium and ytterbium-doped titania thin films deposited by MOCVD}
\date{\today}
\author{Sébastien Forissier}
\email{sebastien.forissier@grenoble-inp.org}
\author{Hervé Roussel}
\author{Carmen Jimenez}
\author{Odette Chaix}
\affiliation{Laboratoire des Matériaux et du Génie Physique, CNRS, Grenoble Institute of Technology, MINATEC, 3 parvis Louis Néel, 38016 Grenoble, France}
\author{Antonio Pereira}
\author{Amina Bensalah-Ledoux}
\affiliation{Laboratoire de Physico-Chimie des Matériaux Luminescents UMR 5620 CNRS / UCBL Domaine Scientifique de la Doua, Université Claude Bernard Lyon 1 Bâtiment Alfred Kastler 10 rue Ada Byron 69622 Villeurbanne cedex, France}
\author{Jean-Luc Deschanvres}
\affiliation{Laboratoire des Matériaux et du Génie Physique, CNRS, Grenoble Institute of Technology, MINATEC, 3 parvis Louis Néel, 38016 Grenoble, France}
\author{Bernard Moine}
\affiliation{Laboratoire de Physico-Chimie des Matériaux Luminescents UMR 5620 CNRS / UCBL Domaine Scientifique de la Doua, Université Claude Bernard Lyon 1 Bâtiment Alfred Kastler 10 rue Ada Byron 69622 Villeurbanne cedex, France}

\begin{abstract}
In this study we synthesized thin films of titanium oxide doped with thulium and/or ytterbium to modify the incident spectrum on the solar cells. This could be achieved either by photoluminescence up-converting devices, or down-converting devices. As down-converter thin films our work deals with thulium and ytterbium-doped titanium dioxide. Thulium and ytterbium will act as sensitizer and emitter, respectively. The rare-earth doped thin films are deposited by aerosol-assisted MOCVD using organo-metallic precursors such as titanium dioxide acetylacetonate, thulium and ytterbium tetramethylheptanedionate solved in different solvents. These films have been deposited on silicon substrates under different deposition conditions (temperature and dopant concentrations for example). Adherent films have been obtained for deposition temperatures ranging from 300\textdegree C to 600\textdegree C. The deposition rate varies from 0.1 to 1 $\upmu$m/h. The anatase phase is obtained at substrate temperature above 400\textdegree C. Further annealing is required to exhibit luminescence and eliminate organic remnants of the precursors. The physicochemical and luminescent properties of the deposited films were analyzed versus the different deposition parameters and annealing conditions. We showed that absorbed light in the near-UV blue range is re-emitted by the ytterbium at 980~nm and by a thulium band around 800~nm.
\end{abstract}


\maketitle

\section{Introduction}
Titanium dioxide has received much interest because of its various applications: photocatalysis, pigment, Transparent Conducting Oxide for photovoltaic applications~\cite{Richards2006}. In this field we focused on doping titanium dioxide with rare-earths (thulium and ytterbium in this work) to study the down-conversion between those ions from blue-visible and near-ultraviolet light to near-infrared light for solar spectrum engineering. The modification of the solar spectrum would allow us to achieve better yields for silicon solar cells~\cite{Badescu2007,Trupke2002}. Thin films were grown using aerosol assisted Metal Oxide Chemical Vapor Deposition (MOCVD) method.

\section{Experimental}
Thulium and ytterbium-doped titanium oxide thin films were deposited by mean of aerosol assisted Metal Oxide Chemical Vapor Deposition method~\cite{Deschanvres1998,Deschanvres1989}. Liquid source solution, were composed of titanium (IV) oxide bis(acetylacetonate), TiO$_2$(acac)$_2$, thulium (III) tris(2,2,6,6-tetramethyl-3,5-heptanedionate) and ytterbium (III)(acac) dissolved in high purity butanol (99\%). All these precursors
were purchased from STREM Chemicals, butanol was purchased from Alfa Aesar. Precursors were selected for non-toxicity, good stability at room temperature, easy handling, high volatility and low cost~\cite{Ryabova1968}. The films were deposited on (100) silicon substrate. The source solution is delivered to the piezoelectric transducer~\cite{Viguie1975} through a constant level burette to ensure a constant pulverization during the whole deposition. The aerosol was produced by means of a flat piezoelectric transducer excited at 800~kHz which generated an ultrasonic beam in a solution containing the reactant of the material to be deposited. This ultrasonic spraying system guarantees a narrow dispersion of the droplet size (4-10 $\upmu$m). Droplets are carried with two air fluxes, dried and purified, to the heated sample holder, the lower air flux (12.7~l/m) assure the main propulsion when the upper air flux (10.1~l/m) lengthen the vapour stay in the vicinity of the sample holder. The overall air flow is parallel to the substrate's surface. For the deposition, the substrate was fixed by clips on the sample holder heated by an electrical resistance with a built-in thermocouple.

We made different series of samples with varying cation concentration in solution. Both Tm-doped, Yb-doped and co-doped samples were successfully synthesized. The compositions of the doped films were measured by electron probe microanalysis (EPMA) and computed by help of special software dedicated to the thin film analysis, called Stratagem and edited by the SAMx society~\cite{Pouchou1984}. The properties of the films reported in this paper are described in section 3.2 and summarized in table 1. The X-ray diffraction profile was obtained with a Bruker D8 Advance using Cu K$\alpha_1$ radiation in $\theta/2\theta$ configuration. Absorption FT-IR spectroscopy was used to study the structural evolution of the films
versus the deposition conditions. Spectra were obtained between 250 and 4000 cm$^-1$ with 4 cm$^-1$ resolution with a Bio-Rad Infrared Fourier Transform spectrometer FTS165 and after performing Si substrate subtraction. Fluorescence and decay time measurements were taken using a laboratory-built apparatus. It is composed of a tunable laser NT342-10-AW from Ekspla, a monochromator TRIAX 190 from Jobin Yvon, a photomultiplier C4877 from Hamamatsu and a multichannel scaler SR430 from Stanford Research. Excitation spectra were taken using a F900 spectrofluorimeter Edinburgh with a high spectral resolution. A Xenon Arc lamp (450 W) is used for the excitation, the detector is a photomultiplier Hamamatsu R2658P cooled by Peltier effect.

\section{Results and discussion}
\subsection{Structural properties}
\begin{figure}[ht]
	\includegraphics[width=0.47\textwidth]{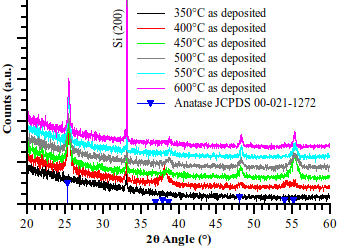}
	\caption{Crystallization versus deposition temperature.}
	\centering
	\label{drx_bruts}
\end{figure}
\begin{figure}[ht]
	\includegraphics[width=0.47\textwidth]{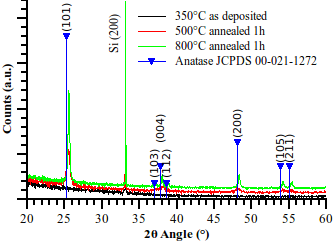}
	\caption{Crystallization effect of annealing.}
	\centering
	\label{drx_recuit}
\end{figure}

As shown by XRD diffraction the samples deposited at temperature lower than 400\textdegree C are amorphous (Figure~\ref{drx_bruts}). Above 400\textdegree C as deposition temperature, the films crystallize in the anatase phase of titanium oxide. As shown on Figure~\ref{drx_recuit} the crystalline quality of the films is improved by annealing, first at 500\textdegree C for 1~h and then at 800\textdegree C for 1~h. With increasing annealing temperature the anatase peaks are becoming higher and thinner. After annealing at 800\textdegree C undoped samples crystallize in the rutile phase, whereas the presence of rare-earth dopants prevents the phase transition as reported in~\cite{Graf2007}.

\subsection{Composition}
\begin{figure}[ht]
	\includegraphics[width=0.47\textwidth]{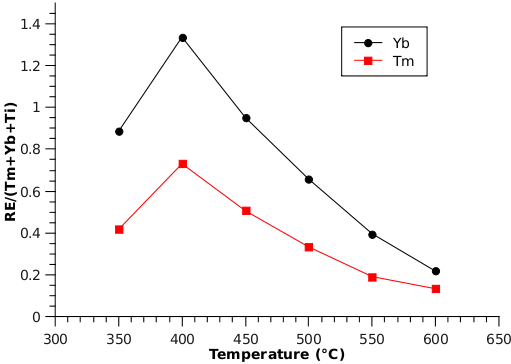}
	\centering
	\caption{Doping versus temperature.}
	\label{dopage}
\end{figure}

Due to the reactor's geometry, the thicknesses of the sample are not homogeneous on the whole surface but it shows a good uniformity on half of it as seen with interferential colors depending on their thickness. However despite this geometry the doping level is quite uniform on the whole surface. We synthesized different samples from the same solution at different temperature and the electron microprobe measurement showed that the optimal doping efficiency is obtained at 400\textdegree C (Figure~\ref{dopage}). In this condition the rare-earth precursor reactivity is lower than the titanium oxide precursor reactivity.

\subsection{Luminescence properties}
\begin{figure}[ht]
	\includegraphics{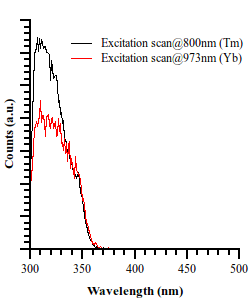}
	\centering
	\caption{Excitation spectrum of a Tm-doped sample (red curve) and excitation spectrum of a Yb-doped sample (black curve). Spectra were recorded in separate experiments.}
	\label{excitation}
\end{figure}
\begin{figure}[ht]
	\includegraphics{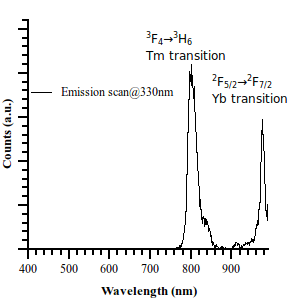}
	\centering
	\caption{Emission spectrum of a Tm, Yb co-doped sample.}
	\label{emission}
\end{figure}

The emission scan (Figure~\ref{emission}) were taken exciting at 330~nm in the TiO$_2$ matrix (3.2 eV~gap) near the$^1D_2$ level of thulium. On a Tm, Yb co-doped sample we see both the Yb $^2F_{5/2}-^2F_{7/2}$ transition at 980~nm and the Tm $^3F_4-^3H_6$ transition around 800~nm. On mono-doped samples (not figured) we only see the respective ion transitions. It appears that the absorbed energy is transferred through the matrix to the rare-earth dopant ions leading to a down-conversion mechanism with thulium and ytterbium. The Yb luminescence is interesting for solar cells because it's right before the band gap of the silicon; the Tm luminescence is still in the wavelength range of good absorption for silicon solar cells.

The excitations scans (Figure~\ref{excitation}) were taken looking at the ytterbium $^2F_{5/2}-^2F_{7/2}$ transition and then the thulium $^3F_4-^3H_6$ transition. Both transitions are activated from the near-UV region between 300 and 350~nm. This corroborates the idea of the energy path we have between the matrix and the rare-earth ions.

\begin{figure}[ht]
	\includegraphics[width=0.47\textwidth]{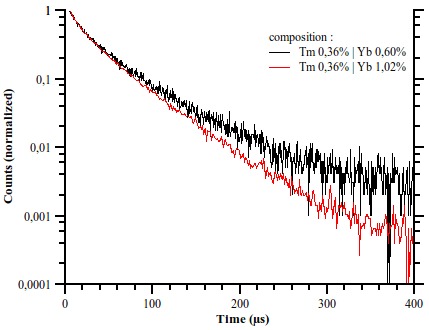}
	\centering
	\caption{Decay time of Tm,Yb-doped samples.}
	\label{declin}
\end{figure}

The lifetime of the $^3F_4-^3H_6$ transition of Tm was measured on Tm-doped only and Tm, Yb co-doped samples. For Tm-doped only samples we see that the lifetime of the transition decreases with the increasing percentage of Tm (see Table~\ref{table}). When we compare co-doped and mono-doped samples we see
that at a constant level of Tm when we increase the percentage of Yb the lifetime is decreasing (Figure~\ref{declin}), leading us to conclude of energy transfer between Tm and Yb. The transfer rate was computed with the relation $1-\tau_x/\tau$ where $\tau_x$ is the lifetime of a co-doped sample and $\tau$ the lifetime of a Tm-doped sample with the same concentration, with samples A and E this gives a 10\% transfer rate.

\begin{table}
\renewcommand{\arraystretch}{1.2}
\vspace*{2ex}\begin{tabular}{cccc}
\hline\hline Sample & Tm (\%) & Yb (\%) & Lifetime ($\upmu$) \\ 
A & 0.86 & 0 & 50 \\ 
B & 2.08 & 0 & 48 \\ 
C & 0.41 & 0.69 & 56 \\ 
D & 0.37 & 0.85 & 46 \\ 
E & 0.83 & 0.79 & 45 \\ 
F & 1.21 & 3.63 & 43 \\ 
\hline\hline
\end{tabular} 

\caption{Cationic doping and lifetimes of the thulium}
\label{table}
\end{table}

\section{Conclusion}
In summary, we succeeded in doping titanium dioxide with thulium and ytterbium by Metal Oxide Chemical Vapour and to grow them partially crystallized. After proper annealing the luminescence of the samples was studied both with excitation and emission scans. The luminescence study showed energy transfer between the titanium oxide matrix and the rare-earth ions and two interesting emission for silicon solar cells. Lifetime measurements showed the energy exchange between thulium and ytterbium with an estimated transfer rate of 10\%.

\section{Acknowledgements}
This work has been supported by French Research National Agency (ANR) through Habitat intelligent et solaire photovoltaïque program (project MULTIPHOT n\textdegree ANR-09-HABISOL-009), the CARNOT Energie du future and the cluster Energies Rhône-Alpes.

\bibliographystyle{unsrt}
\bibliography{/home/forissis/these/biblio/biblio}

\begin{thebibliography}{1}

\bibitem{Richards2006}
BS~Richards.
\newblock Luminescent layers for enhanced silicon solar cell performance:
  Down-conversion.
\newblock {\em Solar energy materials and solar cells}, 90(9):1189--1207, may
  2006.

\bibitem{Badescu2007}
Viorel Badescu, Alexis De~Vos, Alina~Mihaela Badescu, and Aleksandra Szymanska.
\newblock Improved model for solar cells with down-conversion and down-shifting
  of high-energy photons.
\newblock {\em Journal of physics D-applied physics}, 40(2):341--352, JAN 21
  2007.

\bibitem{Trupke2002}
T~Trupke, MA~Green, and P~Wurfel.
\newblock Improving solar cell efficiencies by up-conversion of sub-band-gap
  light.
\newblock {\em J. Appl. Phys.}, 92(7):4117--4122, OCT 1 2002.

\bibitem{Deschanvres1998}
J.~L. Deschanvres, W.~Meffre, J.~C. Joubert, J.~P. Senateur, F.~Robaut, J.~E.
  Broquin, and R.~Rimet.
\newblock Rare earth-doped alumina thin films deposited by liquid source cvd
  processes.
\newblock {\em J. Alloys Compd.}, 275-277:742 -- 745, 1998.

\bibitem{Deschanvres1989}
J-L Deschanvres, F~Cellier, G~Delabouglise, M~Labeau, M~Langlet, and J-C
  Joubert.
\newblock {Thin-film of ceramic oxides by modified CVD}.
\newblock {\em {Journal de physique}}, {50}({C-5, 5}):{695--705}, May {1989}.

\bibitem{Ryabova1968}
L.A. Ryabova and Y.~S. Savitskaya.
\newblock The preparation of thin films of some oxides by the pyrolysis method.
\newblock {\em Thin Solid Films}, 2:141, 1968.

\bibitem{Viguie1975}
J.C. Viguié and J~Spitz.
\newblock Chemical vapor deposition at low temperatures.
\newblock {\em J. Electrochem. Soc.}, 122(4):585, 1975.

\bibitem{Pouchou1984}
J.~L. Pouchou and F.~Pichoir.
\newblock {\em La Recherche Aérospatiale}, 5:349--367, 1984.

\bibitem{Graf2007}
Corinna Graf, Renate Ohser-Wiedemann, and Guenter Kreisel.
\newblock Preparation and characterization of doped metal supported
  tio2-layers.
\newblock {\em Journal of photochemistry and photobiology A-chemistry},
  188(2-3):226--234, MAY 20 2007.

\end{thebibliography}
\end{document}